\documentclass[submission,copyright,creativecommons]{eptcs}
\providecommand{\event}{MCU 2013} % Name of the event you are submitting to
\usepackage{breakurl}             % Not needed if you use pdflatex only.
\usepackage{amsmath,amssymb}
\usepackage{xspace}
\usepackage{graphicx}
\usepackage{amsthm}
\usepackage[capitalize]{cleveref} %automatically add the name of the object referred to
\newcommand{\lab}[1]{\label{#1}\marginpar{\footnotesize #1}} 
\renewcommand{\lab}[1]{\label{#1}}
\makeatletter
\DeclareRobustCommand*{\bfseries}{%
  \not@math@alphabet\bfseries\mathbf
  \fontseries\bfdefault\selectfont
  \boldmath
}   
\makeatother

\newcommand{\IR}{\mathbb{R}}
\newcommand{\IP}{\mathbb{P}}
\newcommand{\IPR}{\IP^2(\IR)}
\newcommand{\IPF}{\IP^2(\IF)}
\newcommand{\IPQ}{\IP^2(\IQ)}

\newcommand{\IN}{\mathbb{N}}
\newcommand{\IQ}{\mathbb{Q}}
\newcommand{\IF}{\mathbb{F}}
\newcommand{\BSS}{BSS\xspace}
\newcommand{\Ring}{R}

\newcommand{\calR}{\mathcal{R}}
\newcommand{\calP}{\textbf{\textup{P}}}
\newcommand{\calRP}{\textbf{\textup{RP}}}
\newcommand{\calNP}{\textbf{\textup{NP}}}
\newcommand{\calNPR}{\calNP_{\!\Ring}}
\newcommand{\calNPIR}{\calNP_{\!\IR}}
\newcommand{\calNPIF}{\calNP_{\!\IF}}
\newcommand{\PSPACE}{\textbf{\textup{PSPACE}}}
\newcommand{\calBP}{\operatorname{\textbf{\textup{BP}}}}
\newcommand{\calPSPACE}{\PSPACE}

\newcommand{\QSAT}{\operatorname{\textup{\sf QSAT}}}

\newcommand{\FEAS}{\operatorname{\textup{\sf FEAS}}}
\newcommand{\XSAT}{\operatorname{\textup{\sf XSAT}}}
\newcommand{\UNITVEC}{\operatorname{\textup{\sf XUVEC\,}}}
\newcommand{\NONTRIV}{\operatorname{\textup{\sf XNONTRIV\,}}}
\newcommand{\NONEQUIV}{\operatorname{\textup{\sf XNONEQUIV\,}}}

\newcommand{\aname}[1]{\textsf{#1}}
\newcommand{\person}[1]{\textsf{#1}}
\newcommand{\id}{\operatorname{id}}
\newcommand{\adjoint}{\dagger}
\newcommand{\sdzero}{\textup{\texttt{0}}\xspace}
\newcommand{\sdone}{\textup{\texttt{1}}\xspace}
\newcommand{\mycite}[2]{{\rm\cite[\textsc{#1}]{#2}}}
\newtheorem{theorem}{Theorem}
\newtheorem{fact}[theorem]{Fact}
\newtheorem{example}[theorem]{Example}
\newtheorem{remark}[theorem]{Remark}
\newtheorem{definition}[theorem]{Definition}
\newtheorem{lemma}[theorem]{Lemma}

\newtheorem{observation}[theorem]{Observation}

\providecommand{\doi}[1]{\href{http://dx.doi.org/#1}{\texttt{doi:#1}}}

\title{Satisfiability of cross product terms is complete for real
nondeterministic polytime Blum-Shub-Smale machines\thanks{%
Supported in parts by the
\emph{Marie Curie International Research
Staff Exchange Scheme Fellowship} \texttt{294962}
within the 7th European Community Framework Programme}}
\author{Christian Herrmann
\and Johanna Sokoli
\institute{\makebox[0pt]{Dept. of Mathematics, TU Darmstadt, GERMANY}}
\and Martin Ziegler
}
\def\titlerunning{Satisfiability over crossproducts is $\calNPIR$--complete}
\def\authorrunning{C. Herrmann, J. Sokoli, M. Ziegler}
\begin{document}
\maketitle
\stepcounter{footnote}

\begin{abstract}
Nondeterministic polynomial-time Blum-Shub-Smale Machines over the reals
give rise to a discrete complexity class between $\calNP$ and $\calPSPACE$.
Several problems, mostly from real algebraic geometry / polynomial systems, 
have been shown complete (under many-one reduction
by polynomial-time Turing machines) for this class.
We exhibit a new one based on questions about 
expressions built from cross products only.
\end{abstract}
%

%%%%%%%%%%%%%%%%%%%%%%%%%%%%%%%%%%%%%%%%%%%%%%%%%%%%%%%%%%%%%%

\section{Motivation}
The Millennium Question ``$\calP$ vs. $\calNP$'' 
asks whether polynomial-time algorithms that may 
guess, and then verify, bits can be turned into deterministic ones.
It arose from the \aname{Cook--Levin--Theorem} asserting
Boolean Satisfiability to be complete for $\calNP$;
which initiated the identification of more and more
other natural problems also complete \cite{GareyJohnson}.

The Millennium Question is posed  \cite{Smale} also for 
models able to guess objects more general than bits. 
More precisely a Blum-Shub-Smale (\BSS) machine 
over a ring $\Ring$ may operate on elements from
$\Ring$ within unit time. It induces the nondeterministic polynomial-time 
complexity class $\calNPR$; for which the following problem 
$\FEAS_\Ring$ 
has been shown complete \mycite{Main Theorem}{BSS}:
\begin{quote} \it
Given\footnote{e.g. as lists of monomials and their coefficients
or as algebraic expressions} a system of multivariate polynomials over $\Ring$, \\
does it admit a joint root from $\Ring$ ?
\end{quote}
See also \mycite{Theorem~3.1}{Cucker} or \mycite{\S5.4}{BCSS}. 
More precisely 
$\FEAS_{\Ring}\subseteq\Ring^*$ is $\calNPR$--complete
with respect to many-one (aka Karp) reducibility by 
polynomial-time \BSS-machines \emph{with} the capability
to peruse finitely many fixed constants from $\Ring$.
\BSS Machines with\emph{out} constants on the other hand
give, restricted to \emph{binary} inputs, rise to the 
discrete complexity class $\calBP(\calNPR^0)$
\mycite{Definition~3.2}{Meer}; for which the following 
problem $\FEAS^0_{\Ring}\subseteq\{\sdzero,\sdone\}^*$ 
is complete under many-one reduction by polynomial-time 
Turing machines:
\begin{quote} \it
Given a system of multivariate polynomials with $0$s and $\pm1$s as coefficients, \\
does it admit a joint root from $\Ring$ ?
\end{quote}
\BSS machines over $\IR$
coincide with the real-RAM model from \aname{Computational Geometry}
\cite{compGeom} and underlie algorithms in \aname{Semialgebraic Geometry}
\cite{Giusti,Lecerf,Scheiblechner1}. 
They give rise to a particularly rich structural complexity theory 
resembling the classical Turing Machine-based one
-- but often (unavoidably) with surprisingly different proofs
\cite{Buergisser2,PCP}.
It is known that $\calNP\subseteq\calBP(\calNPIR^0)\subseteq\calPSPACE$
holds \cite{Grigoriev,Canny,Heintz,Renegar2}.
$\FEAS_{\IR}$ and $\FEAS_{\IR}^0$ 
are sometimes referred to as \aname{existential theory over the reals}.
However even in this highly important case $\Ring=\IR$,
and in striking contrast to $\calNP$, 
relatively few other natural problems  
have yet been identified as complete:
\begin{itemize}
\item Several questions about systems of polynomials \cite{Rossello2,Koiran99}
\item Stretchability of pseudoline arrangements \cite{Shor}
\item Realizability of oriented matroids \cite{Richter}
\item Loading neural networks with real weights \cite{Zhang}
\item Several geometric properties of graphs \cite{Schaefer2}
\item Satisfiability in Quantum Logic $\QSAT$, starting from dimension 3 \cite{LiCS}.
\end{itemize}
The present work extends this list:
We study questions about expressions
built using variables and the cross (aka vector) product ``$\times$''
only, and we establish some of them complete for $\calNPIR$
or $\calBP(\calNPIR^0)$.
These problems are in a sense `simplest' as they involve
only one binary operation symbol (as opposed to
$+,\cdot$ for $\FEAS^0_{\IR}$ or $\vee,\neg$ for $\QSAT$);
in fact so simple that their (trans-$\calNP$) hardness may appear as surprising.

\begin{remark} \lab{r:Identity}
Another decision problem related to $\FEAS_{\Ring}$ and $\FEAS^0_{\Ring}$ is %\aname{Identity Testing}:
the question of whether a given multivariate polynomial $p$ is identically zero or not.
In dense representation (list of monomials and coefficients)
this can easily be solved (over rings $\calR$ of characteristic 0) 
by checking whether all coefficients vanish or not. However when $p$ is given
as a expression, expanding that based on the distributive law may result in an
exponential blow-up of description length. The following \aname{Polynomial Identity Testing} problem
is thus not known to be polytime decidable:
\begin{quote} \it
Given a multivariate ring term $p(X_1,\ldots,X_n)$ with constants $0$ and $\pm1$, \\
does it admit an assignment $x_1,\ldots,x_n$ such that
$p(x_1,\ldots,x_n)\neq0$ \end{quote}
It can be solved, though, in randomized polytime with one-sided error
(class $\calRP\subseteq \calNP$) based on the 
\aname{Schwartz-Zippel Lemma}, cmp. 
{\rm\cite[\S1.5 and \textsc{Thm~7.2}]{Motvani}}.
\end{remark}

%%%%%%%%%%%%%%%%%%%%%%%%%%%%%%%%%%%%%%%%%%%%%%%%%%%%%%%%%%%%%%
\section{Cross Product and Induced Problems}
The cross product in $\IR^3$ is well-known due to its many
applications in physics such as torque or electromagnetism.
Mathematically it constitutes the mapping
\begin{equation} \label{e:CrossProd}
\times : \IR^3\times\IR^3 \;\ni\; \big((v_0,v_1,v_2),(w_0,w_1,w_2)\big)
\;\mapsto\; (v_1w_2-v_2w_1,v_2w_0-v_0w_2,v_0w_1-v_1w_0) \;\in\;\IR^3 \enspace .
\end{equation}
It is bilinear (thus justifying the name ``product'')
but anti-commutative 
$\vec v\times\vec w=-\vec v\times\vec w$ and non-associative
and fails the cancellation law. The following is easily 
verified:

\begin{fact} \lab{f:Tools}
\begin{enumerate}
\itemsep0pt
\item[a)] 
For any independent $\vec v, \vec w$,
the cross  product $\vec u =\vec v \times \vec w$ 
is uniquely determined by   the following:
$\vec u\bot \vec v,\;\vec u \bot \vec w $
(where ``$\bot$'' denotes orthogonality), the triplet $\vec v, \vec w ,\vec u$
is right-handed,   and  lengths  satisfy $\|\vec u\| =\|\vec v\|\cdot \|\vec w\|
\cos \angle(\vec v,\vec w) $. 
In particular, parallel $\vec v,\vec w$ are mapped to $\vec 0$.
\item[b)]
Cross products commute with simultaneous  
orientation preserving orthogonal transformations: 
For $O\in\IR^{3\times 3}$ with $O\cdot O^\adjoint=\id$ and $\det(O)=1$
it holds
$(O\cdot\vec v)\times(O\cdot\vec w)=O\cdot(\vec v\times\vec w)$,
where $O^\adjoint$ denotes the transposed matrix.
\end{enumerate}
\end{fact}

\begin{definition} \lab{d:Syntax}
Fix a field $\IF\subseteq\IR$.
\begin{enumerate}
\itemsep0pt
\item[a)] A \aname{term} $t(V_1,\ldots,V_n)$ 
  (over ``$\times$'', in variables $V_1,\ldots,V_n$)
  is either 
  one of the variables or $(s\times t)$ for terms $s,t$ 
  (in variables $V_1,\ldots,V_n$).
\item[b)] For $\vec v_1,\ldots,\vec v_n\in\IF^3$ the
  \aname{value} $t(v_1,\ldots,v_n)$ is defined inductively
  via \cref{e:CrossProd}.
\item[c)] A term \aname{with affine constants} is a term 
  $t(V_1,\ldots,V_n;W_1,\ldots,W_m)$
  where variables $W_1,\ldots,W_m$ 
  have been pre-assigned certain values $\vec w_1,\ldots,\vec w_m\in\IR^3$.
\item[d)] Recall that 
  $\IPF:=\{\ \IF\vec v:\vec0\neq\vec v\in\IF^3\}$
  denotes the  real  projective plane, where $\IF\vec v=\{\lambda\vec v:\lambda\in\IF\}$.
  For distinct $\IF\vec v,\IF\vec w\in\IPF$ 
  (well-)define $(\IF\vec v)\times(\IF\vec w):=\IF(\vec v\times\vec w)$;
  $\IF\vec v\times\IF\vec v$ is undefined.
\item[e)]
  For a term $t(V_1,\ldots,V_n)$ and $\IF\vec v_1,\ldots,\IF\vec v_n\in\IPF$,
  the \aname{value} $t(\IF\vec v_1,\ldots,\IF\vec v_n)$ is defined inductively via d),
  provided all sub-terms are defined.
\item[f)]
  A term \aname{with projective constants} is a term
  $t(V_1,\ldots,V_n;W_1,\ldots,W_m)$
  where variables $W_1,\ldots,W_m$ 
  have been pre-assigned certain values $\IR\vec w_1,\ldots,\IR\vec w_m\in\IPR$.
\end{enumerate}
\end{definition}
\noindent
Note that every term admits an affine assignment making it evaluate to $\vec 0$.
Some terms in fact always evaluate to $\vec 0$; 
equivalently: are projectively undefined everywhere.

\begin{example} \lab{x:Zero}
Consider the term $t(V,W):=\Big(\big(V\times(V\times W)\big)\times V\Big)
\times(V\times W)$. 
Observe that $\vec v$, $\vec v\times\vec w$, and $\vec v\times(\vec v\times\vec w)$ together
form an orthogonal system for any non-parallel $\vec v,\vec w$.
Moreover $\big(\vec v\times(\vec v\times\vec w)\big)\times\vec v$ is parallel
to $\vec v\times\vec w$.
Therefore $t(\vec v,\vec w)=\vec 0$ holds for every choice of $\vec v,\vec w\in\IR^3$.
\end{example}
\noindent
We are interested in the computational complexity 
of the following discrete decision problems:

\begin{definition} \lab{d:Problems}
\begin{enumerate}
\itemsep0pt
\item[a)] $\NONTRIV^0_{\IF^3}\;:=\;\big\{ \langle t(V_1,\ldots,V_n)\rangle 
  \;\big|\; n\in\IN, \; \exists \vec v_1,\ldots,\vec v_n\in\IF^3: 
  t(\vec v_1,\ldots,\vec v_n)\neq\vec 0\big\}$.
%  \;\subseteq\;\{\sdzero,\sdone\}^*$.
\item[b)] $\NONTRIV^0_{\IPF}\;:=\;\big\{ \langle t(V_1,\ldots,V_n)\rangle
  \;\big|\; n\in\IN, \;\exists \IF\vec v_1,\ldots,\IF\vec v_n]\in\IPF: 
   t(\IF\vec v_1,\ldots,\IF\vec v_n) \text{ defined}\big\}$.
\item[c)] $\UNITVEC^0_{\IF^3}\;:=\;\big\{ \langle t(V_1,\ldots,V_n)\rangle 
  \;\big|\; n\in\IN, \; \exists \vec v_1,\ldots,\vec v_n\in\IF^3: 
  t(\vec v_1,\ldots,\vec v_n)=\vec e_3:=(0,0,1)\big\}$.
\item[d)] $\NONEQUIV^0_{\IPF}\;:=\;\big\{ \langle s(V_1,\ldots,V_n),t(V_1,\ldots,V_n)\rangle 
  \;\big|\;  
\\ \hspace*{\fill}
n\in\IN, \;\exists \IF\vec v_1,\ldots,\IF\vec v_n\in\IPF: 
  s(\IF\vec v_1,\ldots,\IF\vec v_n)\neq t(\IF\vec v_1,\ldots,\IF\vec v_n)
  \text{, both sides defined}\big\}$.
\item[e)] $\XSAT^0_{\IF^3}\;:=\;\big\{ \langle t_1(V_1,\ldots,V_n) \rangle 
  \;\big|\; 
n\in\IN, \;\exists \vec v_1,\ldots,\vec v_n\in\IF^3: t(\vec v_1,\ldots,\vec v_n)=\vec v_1\neq\vec 0
   \big\}$.
\item[f)] $\XSAT^0_{\IPF}\;:=\;\big\{ \langle t_1(V_1,\ldots,V_n) \rangle 
  \;\big|\; 
n\in\IN, \;\exists \IF\vec v_1,\ldots,\IF\vec v_n\in\IPF: t(\IF\vec v_1,\ldots,\IF\vec v_n)=\IF\vec v_1 \big\}$.
\end{enumerate}
\noindent
\emph{Real} variants of problems a) to f) with\emph{out} superscript $0$ are 
defined similarly for input terms \emph{with} constants; e.g.
$\XSAT_{\IR^3}\;:=\;\big\{ \langle t_1(V_1,\ldots,V_n;\vec w_1,\ldots,\vec w_k) \rangle 
  \;\big|\; n,k\in\IN,  \;\vec w_1,\ldots,\vec w_k\in\IR^3\; 
\\ \hspace*{\fill} \exists \vec v_1,\ldots,\vec v_n\in\IR^3: 
  t(\vec v_1,\ldots,\vec v_n;\vec w_1,\ldots,\vec w_k)=\vec v_1\neq\vec 0
   \big\}\;\subseteq\;\IR^*$.
\end{definition}
\noindent
Our main result is

\begin{theorem} \lab{t:Main}
\begin{enumerate}
\item[a)] Among the above discrete decision problems,
  $\NONTRIV^0_{\IR^3}$, $\NONTRIV^0_{\IPR}$, $\UNITVEC^0_{\IR^3}$, and $\NONEQUIV^0_{\IPR}$
  are polytime equivalent to polynomial identity testing (and in particular belong to $\calRP$).
\item[b)] For any fixed field $\IF\subseteq\IR$, the discrete decision problems
  $\XSAT^0_{\IF^3}$ and $\XSAT^0_{\IPF}$ are $\calBP(\calNPIF^0)$--complete.
\item[c)] $\XSAT_{\IR^3}$ and $\XSAT_{\IPR}$ are $\calNPIR$--complete.
\end{enumerate}
\end{theorem}
\medskip
\noindent
This establishes a normal form for cross product equations
with a variable on the right-hand side ---
in spite of the lack of a cancellation law.

%%%%%%%%%%%%%%%%%%%%%%%%%%%%%%%%%%%%%%%%%%%%%%%%%%%%%%%%%%%%%%%%%%%%%%%%%%%%%%%%
\section{Proofs}
$\NONTRIV^0_{\IPF}$ is equal to $\NONTRIV^0_{\IF^3}$ as a set;
and it holds $\NONTRIV^0_{\IPR}=\UNITVEC^0_{\IR^3}$:
Suppose $t(\vec v_1,\ldots,\vec v_n)=:\vec w\neq\vec 0$.
Since $t$ is homogeneous in each coordinate, 
by suitably scaling some argument $\vec v_j$ 
we may w.l.o.g. suppose\footnote{This requires taking square roots} $|\vec w|=1$. 
Now take an orientation preserving orthogonal transformation $O$ with $O\cdot\vec w=\vec e_3$:
\Cref{f:Tools}b) yields $t(O\cdot\vec v_1,\ldots,O\cdot\vec v_n)=\vec e_3$.
Concerning the reduction from $\NONEQUIV^0_{\IPF}$ to $\NONTRIV^0_{\IF}$ observe that,
for $\vec v_1,\ldots,\vec v_n\in\IF^3\setminus\{\vec 0\}$,
$\IF s(\vec v_1,\ldots,\vec v_n)\neq \IF t(\vec v_1,\ldots,\vec v_n)$
implies $s(\vec v_1,\ldots,\vec v_n)\times t(\vec v_1,\ldots,\vec v_n)\neq0$
and vice versa. Conversely an instance to $\NONTRIV^0_{\IF}$ is 
either a variable (trivial case) or of the form $s\times t$;
in which case nontriviality is equivalent to projective nonequivalence
of $s,t$.

\medskip
We now reduce $\NONTRIV^0_{\IR^3}$ to polynomial identity testing, observing
that $\vec u\times\vec v$ is a triple of bilinear polynomials 
in the 6 variables $u_x,u_y,u_z,v_x,v_y,v_z$ with coefficients $0,\pm1$.
Thus, $t(\vec v_1,\ldots,\vec v_n)$ amounts to 
a triple of terms $p_x,p_y,p_z$ in $3n$ variables with coefficients $0,\pm1$.
Now by construction a real assignment $\vec v_1,\ldots,\vec v_n$ makes $t$ evaluate to nonzero 
~iff~ the three terms $p_x,p_y,p_z$ do not simultaneously evaluate to zero.
This yields the reduction $t\mapsto p_x^2+p_y^2+p_z^2$.

\medskip
Concerning $\XSAT_{\IR^3}$, 
a nondeterministic real \BSS machine can, given a term 
$t(V_1,\ldots,V_n;\vec w_1,\ldots,\vec w_k)$ with constants $\vec w_j\in\IR^3$, 
in time polynomial in the length of $t$ guess an assignment $\vec v_1,\ldots,\vec v_n\in\IR^3$
and apply \cref{e:CrossProd} to evaluate $t$ and verify the result to be nonzero.
Similarly a nondeterministic \BSS machine over $\IF$ can,
given a term $t(V_1,\ldots,V_n)$ without constants, 
in polytime guess and evaluate it on an assignment
$\vec v_1,\ldots,\vec v_n\in\IF^3$.

\medskip
$\XSAT^0_{\IPR}$ reduces to $\XSAT^0_{\IR^3}$ in polytime as follows:
For any $\vec w$ non-parallel to $\vec t$,
$\vec t':=(\vec t\times\vec w)\times\big((\vec t\times\vec w)\times t\big)$
is a multiple of $\vec t$; see \cref{f:crossframe}a).
Note that scaling $\vec w$ affects $\vec t'$ quadratically.
Similarly, $\big(\vec w\times(\vec t\times\vec w)\big)\times\vec t$
is a multiple of $\vec t\times\vec w$; and replacing it in the
first subterm defining $\vec t'$ (and renaming $\vec t,\vec t'$ to
$\vec s,\vec s'$) shows that $\vec s':=
\Big(\big(\vec w\times(\vec s\times\vec w)\big)\times\vec s\Big)
\times\big(\vec s\times(\vec s\times\vec w)\big)$
is a multiple of $\vec s$; one scaling cubically with $\vec w$.
So $\IR$ being closed under cubic roots,
$s(V_1,\ldots,V_n)=V_1$ is satisfiable over $\IPR$ ~iff~
$s(V_1,\ldots,V_n)=\lambda^3 V_1$ is satisfiable over $\IR^3$ 
for some $\lambda\in\IR$ ~iff~
$s'(V_1,\ldots,V_n,W)=V_1$ is satisfiable over $\IR^3$, where
$s':=\Big(\big(W\times(s\times W)\big)\times s\Big)\times\big(s\times(s\times W)\big)$.
The reduction for the case \emph{with} constants,
that is from $\XSAT_{\IPR}$ to $\XSAT_{\IR^3}$, works similarly.

%%%%%%%%%%%%%%%%%%%%%%%%%%%%%%%%%%%%%%%%%%%%%%%%%%%%%%%%%%%%%%%%%%%%%%%%%%%%%%%%
\subsection{Hardness}
It remains to reduce (in polynomial time)
\begin{enumerate}
\item[i)] $\FEAS_{\IR}$ to $\XSAT_{\IPR}$ and
\item[ii)] $\FEAS^0_{\IF}$ to $\XSAT^0_{\IPF}$ and
\item[iii)] polynomial identity testing to $\NONTRIV^0_{\IPR}$.
\end{enumerate}
These can be regarded as quantitative refinements of \cite{Havlicek}.
We first recall some elementary, but useful facts
about the cross product in the projective setting.
\begin{fact} \label{f:ele}   Consider $U,V,W,T \in \IPF$. By `plane' we mean 
  $2$-dimensional linear subspace.
\begin{enumerate} 
\item[1)]    $U=V \times W$ ~iff~
the  plane orthogonal to $U$ is spanned by $V,W$.  In particular,
$V \times W=W \times V$.
\item[2)] If   $V \times W$ and $ U\times T$ are defined then
 $(V \times W) \times (U\times T)$  
 is the intersection of the plane spanned  by $V,W$ 
 with the plane spanned by $U,T$;
 undefined if this intersection is degenerate.
\item[3)] 
  $V\times (W\times V)$ is the orthogonal projection of 
$W$ into the plane orthogonal to $V$;
undefined iff $W=V$, i.e. in case the projection is degenerate.
\end{enumerate}
\end{fact} 
\noindent
The following considerations are heavily inspired by 
the works of  \person{John von Neumann} but for the
sake of self-containment here boiled down explicitly.

\begin{lemma} \lab{l:vonStaudt}
Fix a subfield $\IF$ of $\IR$.
Let $\vec v_1,\vec v_2,\vec v_3$ denote an 
ortho\emph{go}nal basis of $\IF^3$. Then  
$V_j:=\IF\vec v_j$ satisfies $V_1\times V_2=V_3$, $V_2\times V_3=V_1$,
and
$V_3\times V_1=V_2$. Moreover abbreviating
$V_{12}:=\IF(\vec v_1-\vec v_2)$ and $V_{23}:=\IF(\vec v_2-\vec v_3)$
and $V_{13}:=\IF(\vec v_1-\vec v_3)$, we have for $r,s\in\IF$:
\begin{enumerate}
\item[a)] $\IF(\vec v_1-rs\vec v_2)\;=\;V_3\times
 \big[\IF(\vec v_3-r\vec v_2)\times\IF(\vec v_1-s\vec v_3)\big]$
\item[b)] $\IF(\vec v_1-s\vec v_3)\;=\;
  V_2\times\big[V_{23}\times\IF(\vec v_1-s\vec v_2)\big]$
\item[c)] $\IF(\vec v_3-r\vec v_2)\;=\;V_1\times
  \big[V_{13}\times\IF(\vec v_1-r\vec v_2)\big]$
\item[d)] $\IF\big(\vec v_1-(r-s)\vec v_2\big)\;=\;V_3\times
  \big[\big([V_{23}\times\IF(\vec v_1-r\vec v_2)]
\times[V_2\times\IF(\vec v_1-s\vec v_3)]\big)\times V_3\big]$
\item[e)] $V_{13} \;=\; V_2\times(V_{12}\times V_{23})$.
\item[f)] 
For $W\in\IPF$, the expression
  $\imath(W):=(W\times V_3)\times\Big(\big((W\times V_3)\times V_3\big)\times V_2\Big)$ 
  is defined precisely when $W=\IF(\vec v_1-r\vec v_2+s\vec v_3)$
  for some  $s\in\IF$ and a unique $r\in\IF$; and in this case
  $\imath(W)=\IF(\vec v_1-r\vec v_2)$. Moreover, if  $W=\IF(\vec v_1
  -r\vec v_2)$ then $\imath(W)=W$.
\end{enumerate}
\end{lemma}
\noindent
Note that the $V_j$ here do not denote variables but elements of
$\IPF$. 
Concerning the proof  of Lemma \cref{l:vonStaudt},
 e.g. for a) 
observe that $\vec v_1 -rs \vec v_2 = \vec v_1 -s\vec v_3 -s(  \vec v_3 -r \vec
v_2)$ is orthogonal to $V_3$ and
contained in the plane spanned by  $\vec v_3 -r \vec v_2$.
In d)  one applies   3) of  \cref{f:ele}  with  subterm $W$ evaluating to 
$\IF(\vec v_1 - (r-s)\vec v_2 -s\vec v_3)$  in  view of 2).
For f) observe that, if $W$ lies in the $V_2$--$V_3$--plane,
its projection $(W\times V_3)\times V_3$ according to 3)
coincides with $V_2$ (corresponding to slope $r=\pm\infty$)
and renders the entire term undefined;
whereas for $W$ not in the $V_2$--$V_3$--plane,
$\big((W\times V_3)\times V_3\big)\times V_2$ coincides with $V_3$.

\medskip
Let us abbreviate $\bar V:=(V_1,V_2,V_3,V_{12},V_{23})$ 
derived from an orthogonal basis $\vec v_1,\vec v_2,\vec v_3$ as above.
In terms of \person{von Staudt}'s encoding of elements $r\in\IF$
as projective points $\Theta_{\bar V}(r):=\IF(\vec v_1-r\vec v_2)\perp\IF\vec v_3$, 
\cref{l:vonStaudt}a+d) demonstrate how to express the ring
operations using only the crossproduct; note that
$r+s=r-(0-s)$ where $0\in\IF$ is encoded as $V_1$.
\Cref{l:vonStaudt}a) involves two other encodings 
such as $\IF(\vec v_1-s\vec v_3)$,
but \cref{l:vonStaudt}b+c) exhibit how to
express these using the cross product and $\Theta_{\bar V}$ only
as well as $V_{23}$ and $V_{13}$.
$V_{13}$ can even be disposed off by means of \cref{l:vonStaudt}e).
Plugging b)+c)+e) into a) and d), we conclude that there exist
cross product terms $\ominus(R,S;\bar V)$ 
and $\otimes(R,S;\bar V)$ 
in variables $R,S$ with constants $\bar V=\big(V_1=\Theta_{\bar V}(0),
V_2,V_3,V_{12}=\Theta_{\bar V}(1),V_{23}\big)$ as above
such that for every $r,s\in\IF$ it holds
$\Theta_{\bar V}(rs)=\otimes\big(\Theta_{\bar V}(r),\Theta_{\bar V}(s);\bar V\big)$ 
and $\Theta_{\bar V}(r-s)=\ominus\big(\Theta_{\bar V}(r),\Theta_{\bar V}(s);\bar V\big)$ 

Now any polynomial $p\in\IF[X_1,\ldots,X_n]$ is composed, using
the two ring operations, 
from variables and coefficients from $\IF$.
More precisely, according to \cref{l:vonStaudt},
the above encoding 
extends to a mapping $\Theta_{\bar V}$ assigning, to any ring term $p(X_1,\ldots,X_n)$ 
with constants $c\in\IF$, some cross product term 
$t_p$ in variables $X_1,\ldots,X_n$ 
with constants  $\Theta_{\bar V}(c) \in \IPF$
and constants $V_1,V_2,V_3,V_{12},V_{23}\in\IPF$; 
moreover $\Theta_{\bar V}$ `commutes' with  the map $p \mapsto t_p$
in the sense that 
\begin{equation} \label{e:Interpret}
t_p\big(\Theta_{\bar V}(x_1),\ldots,\Theta_{\bar V}(x_n)\big) \;=\;
\Theta_{\bar V}\big(p(x_1,\ldots,x_n)\big) \enspace .
\end{equation}
Since $t_p$ is defined by structural induction over $p$
using the constant-size terms from \cref{l:vonStaudt},
it can be evaluated by a \BSS machine in time polynomial
in the description length of the ring term $p$.

Moreover by \cref{l:vonStaudt}f) precisely the $\imath_{\bar V}(W)$ 
are images under $\Theta_{\bar V}$.
Thus, every satisfying assignment to the cross product equation
\begin{equation} \label{e:Interpret2}
t'_p \;\;:=\;\; \Big(t_p\big(\imath(X_1),\ldots,\imath(X_n)\big)\;=\;V_1\Big)
\end{equation}
comes from a root  $(r_1, \ldots ,r_n)$ of $p$;
namely the unique $r_j$ such that $X_j=\IF(\vec v_1+r_j\vec v_2+s_j\vec v_3)$.
Conversely, given a root $(r_1, \ldots ,r_n)$ of $p$,
$X_j:=\Theta_{\bar V}(r_j)$ yields a 
a satisfying assignment for the equation $t'_p=V_1$.
Similarly, (the partial map given by) 
{$t'_p\times V_1$} is nontrivial ~iff~ $p$ is not identically zero.
We have thus proved Claim~i).

In order to establish also the remaining Claims~ii) and iii)
we turn every $d$-variate ring term $p$ with coefficients $0,\pm1$
into an `equivalent' cross product term $t''_p$ with\emph{out} 
constants and in particular avoiding explicit reference to the
fixed $V_1,V_2,V_3,V_{12},V_{23}$ from
\cref{l:vonStaudt} based on the following

\begin{observation} \label{o:Frame}
Fix a subfield $\IF$ of $\IR$. To $A,B,C\in\IPF$ consider
\begin{gather} \label{e:Frame}
V_{12} \;:=\; B \quad
V_2 \;:=\; (A\times B)\times A \quad
V_{23} \;:=\; C\times A \quad
V_1 \;:=\; V_{2}\times V_{23} \quad
V_3 \;:=\; \big(V_{23}\times(B\times V_2)\big)\times B 
\end{gather}
\begin{enumerate}
\item[a)]
These may be undefined in cases such as $A=B$ (whence $V_2=\bot$)
or when $A,C,A\times B$ are collinear (thus $V_{23}=V_2$ and $V_1=\bot$)
or when $A,B,C$ are collinear (where $V_{23}=A\times B$ and $V_3=\bot$)
or when $A\bot B$ (where $B=V_2$ and $V_3=\bot$).
\item[b)]
On the other hand for example $A:=\IF\vec v_1$, 
$B:=\IF(\vec v_2-\vec v_1)$ and $C:=\IF(\vec v_2+\vec v_3)$,
defined in terms of an orthogonal basis,
recover $V_1,V_2,V_3,V_{12},V_{23}$ from \cref{l:vonStaudt}.
\item[c)]
Conversely when all quantities in \cref{e:Frame} are defined,
then $V_1=A$ and there exists a right-handed
orthogonal basis $\vec v_1,\vec v_2,\vec v_3$ of $\IF^3$
such that $V_j=\IF\vec v_j$ and 
$V_{12}=\IF(\vec v_1-\vec v_2)$ and $V_{23}=\IF(\vec v_2-\vec v_3)$.
\end{enumerate}
\end{observation}
\noindent
We may thus replace the tuple of projective constants $\bar V$
in the above reduction $p \mapsto t_p$ mapping a ring term $p(X_1,\ldots,X_n)$ 
to a cross product term $t_p(X_1,\ldots,X_n;\bar V)$
with the subterms $V_1(A,B,C),\ldots,V_{23}(A,B,C)$ 
(considering $A,B,C$ as variables)
according to \cref{o:Frame} to obtain a constant free  cross product term
$t''_p(X_1,\ldots,X_n;A,B,C)$ such that the map $p \mapsto t''_p$ 
commutes with
$\Theta_{\bar V}$ for any projective assignment 
on which $t''_p$ is defined  and  $\bar V(A,B,C)$ 
 given by the  values of the subterms $V_i,V_{ij}$. 

Now let
$\imath(X)$ denote the constant free term from 
\cref{l:vonStaudt}g)  in   variables $X,A,B,C$
(with subterms $V_i$ as above).
Then, from  each  satisfying assignment  to 
$t'''_p:=t''_p\big(\imath(X_1),\ldots,\imath(X_n);A,B,C\big)=A$ 
one obtains as previously again a root  $(r_1, \ldots ,r_n)$ of $p$:
Observation~\ref{o:Frame}c) justifies reusing the reasoning given in
the case with constants. Conversely, given a root 
$(r_1, \ldots ,r_n)$ of $p$ , evaluate $A,B,C$ according
to  Observation \ref{o:Frame}b) and $X_j:=\Theta_{\bar V} (r_j)$
to obtain a satisfying assignment for the equation $t'''_p=A$.
Since the translation $p\mapsto t''_p$ can be carried out
by structural induction in time polynomial in the description
length of $p$, this establishes  Claim~ii).
To deal with iii), consider $t'''_p\times A$.
\qed

\begin{figure}[htb]
\includegraphics[width=\textwidth]{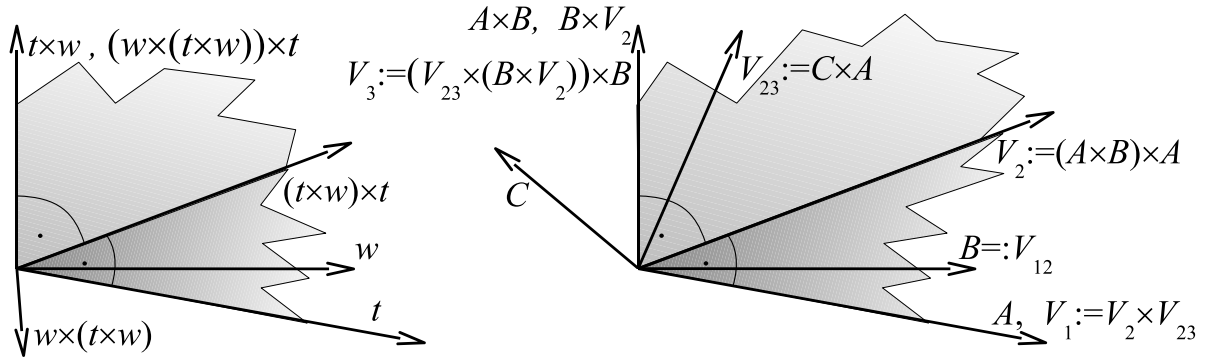}
\caption{\label{f:crossframe}Illustrating the geometry of the terms
considered a) in the reduction from $\XSAT^0_{\IPR}$ to $\XSAT^0_{\IR^3}$
~ and ~ b) in \cref{o:Frame}c.}
\end{figure}
\begin{proof}[Proof of \cref{o:Frame}c)]
By construction, affine lines $A$ and $A\times B$ and $V_2$ 
are pairwise orthogonal; see \cref{f:crossframe}b).
Moreover $A\neq B$ 
because $A\times B$ a subterm of $V_2$ is defined by hypothesis.
Since both $V_2$ and $V_{23}=C\times A$ are orthogonal to $A$,
their projective cross product $V_1$ must coincide with $A$
whenever defined and in particular $V_2\neq V_{23}$;
moreover $V_2$ and $V_{23}$ and $A\times B$ lie in a common plane.
$B\times V_2$ as subterm of $V_3$ being defined
requires $V_2\neq B$; yet these two and $A=V_1$ are orthogonal to
$A\times B$ and thus lie in a common plane.
In particular $B\times V_2=A\times B$.
Finally, $V_{23}$ and $B\times V_2=A\times B$ both
being orthogonal to $A$, their defined cross product
as subterm of $V_3$ requires $V_{23}\neq B\times V_2$
and $V_3=B\times V_2=A\times B$.
To summarize: $V_1,V_2,V_3$ are pairwise orthogonal;
and $V_1,V_{12},V_2$ are pairwise distinct yet all orthogonal to $V_3$;
similarly $V_2,V_{23},V_3$ are pairwise distinct yet all orthogonal to $V_1$.
Now choose $0\neq\vec v_1\in V_1$ arbitrary
and $\vec v_2\in V_2$ such that $V_{12}=\IF(\vec v_1-\vec v_2)$;
finally choose $\vec v_3\in V_3$ such that $V_{23}=\IF(\vec v_2-\vec v_3)$.
If these pairwise orthogonal vectors $\vec v_1,\vec v_2,\vec v_3$ 
happen to form a left-handed system, simply flip all their signs.
\end{proof}
%

%%%%%%%%%%%%%%%%%%%%%%%%%%%%%%%%%%%%%%%%%%%%%%%%%%%%%%%%
\section{Conclusion}

We have 
identified a new problem complete (i.e. universal) for nondeterministic
polynomial-time \BSS machines, namely from exterior algebra:
the satisfiability of a single equation built only by iterating cross products.
This enriches algebraic complexity theory 
and emphasizes the importance of the Turing (!)
complexity class $\calBP(\calNPIR^0)$.

%Not relying on Fact~\ref{f:Tools}c),
%our results extend to the seven-dimensional cross product(s).
Moreover our proof yields a cross product equation
$t'''_{X^2-2}(Y,A,B,C)=A$ 
solvable over $\IPR$ but not over $\IPQ$,
the rational projective plane.
In fact the decidability of $\XSAT^0_{\IPQ}$ is 
equivalent to a long-standing open question \cite{Poonen}.

We wonder about the computational complexity of 
equations over the 7-dimensional cross product.

\nocite{*}
\bibliographystyle{eptcs}

\end{document}